\documentclass[aps,superscriptaddress,showpacs,preprint,pra]{revtex4-1}

\usepackage{epsfig}
\usepackage{amsfonts}
\usepackage{amsmath}
\usepackage{xcolor}
\usepackage[normalem]{ulem}

\begin{document}

\title{Long- and short-range interaction footprints in entanglement entropies of 
two-particle Wigner molecules in 2D quantum traps}

\author{Eloisa Cuestas}
\email{mecuestas@famaf.unc.edu.ar}
\affiliation{Facultad de Matem\'atica, Astronom\'{\i}a y F\'{\i}sica,
Universidad Nacional de C\'ordoba and IFEG-CONICET, Ciudad Universitaria,
X5000HUA C\'ordoba, Argentina}

\author{Mariano Garagiola}
\email{mgaragiola@famaf.unc.edu.ar}
\affiliation{Facultad de Matem\'atica, Astronom\'{\i}a y F\'{\i}sica,
Universidad Nacional de C\'ordoba and IFEG-CONICET, Ciudad Universitaria,
X5000HUA C\'ordoba, Argentina}

\author{Federico~M.~Pont}
\email{pont@famaf.unc.edu.ar}
\affiliation{Facultad de Matem\'atica, Astronom\'{\i}a y F\'{\i}sica,
Universidad Nacional de C\'ordoba and IFEG-CONICET, Ciudad Universitaria,
X5000HUA C\'ordoba, Argentina}

\author{Omar Osenda}
\email{osenda@famaf.unc.edu.ar}
\affiliation{Facultad de Matem\'atica, Astronom\'{\i}a y F\'{\i}sica,
Universidad Nacional de C\'ordoba and IFEG-CONICET, Ciudad Universitaria,
X5000HUA C\'ordoba, Argentina}

\author{Pablo Serra}
\email{serra@famaf.unc.edu.ar}
\affiliation{Facultad de Matem\'atica, Astronom\'{\i}a y F\'{\i}sica,
Universidad Nacional de C\'ordoba and IFEG-CONICET, Ciudad Universitaria,
X5000HUA C\'ordoba, Argentina}

\begin{abstract}
The occupancies and entropic entanglement measures for the ground state of two 
particles in a two-dimensional harmonic anisotropic trap are studied. We 
implement a method to study the large interaction strength limit for different 
short- and long-range interaction potentials that allows to obtain the exact 
entanglement spectrum and several entropies. We show that for long-range 
interactions, the 
von Neumann, min-entropy and the family of R\'enyi entropies remain finite for the 
anisotropic traps and diverge logarithmically for the isotropic traps. 
In the short-range interaction case the entanglement measures diverge for 
any anisotropic parameter due to the divergence of uncertainty in the momentum 
since for short-range interactions the relative position width vanishes.
We also show that when the reduced density matrix has finite support the 
R\'enyi entropies present a non-analytical behaviour. 
\end{abstract}

\maketitle

\section{Introduction}

The physics of systems of confined particles has attracted the 
interest of researchers from many different areas working on both, theoretical and experimental aspects \cite{wigner_1934,wineland_1987}. One feature of such quantum systems that has recently gained 
impulse is the study of entropic measures of entanglement \cite{osenda_2015, garagiola_2016, koscik_2015, lopezrosa_2015,simonovic_2015}. Among the many kind of systems that could be addressed using the physics of the confined systems, in the present work we focus on Wigner molecules, which are the finite-size analogue of Wigner crystals, named after the seminal work of E. Wigner \cite{wigner_1934}. Since the late 80's, when the 
first confined linear chains of ions where reported by D. Wineland \cite{wineland_1987}, there has been an increasing capacity to confine, control and manipulate such entities, and has turned the Wigner molecules into a new platform to test the known, and look for new, traits of quantum 
mechanics \cite{drewsen_2015}. 

Trapped ions are not the only physical systems that allow the 
formation of entities like the Wigner molecules (also named Coulomb crystals \cite{drewsen_2015}). The experimental observation of strongly correlated states in quantum dots has attracted considerable interest \cite{cavaliere_2009}. Wigner molecules have also been observed in two-dimensional semiconductor heteroestructures \cite{andrei_1998,piot_2008}, semiconductor quantum dots \cite{kallikaos_2008}, 
one-dimensional quantum wires \cite{ellenberger_2006,singha_2010,kristinsdottir_2011,meyer_2009}, carbon 
nanotubes \cite{deshpande_2008,pecker_2013}, and in crystalline states 
for dusty plasma \cite{melzer_2003}. Several theoretical 
studies \cite{jauregui_1993,guclu_2008,mendl_2014,gambeta_2014,cavaliere_2014,
cavaliere_2015,cavaliere_2015_2,kylanpaa_2016} have demonstrated that
the physics of these systems with reduced dimensionality usually does not 
depend on the shape of the confinement but on its symmetries and 
strength (see Ref. \cite{pecak_2016} for an example where shape does have an 
influence). In the present work we use a harmonic confinement as a model potential, and represent different physical situations using different interaction potentials. Wigner molecules arise when the interparticle interaction strength is much larger than the kinetic energy. The latter can be related to the temperature of the system \cite{drewsen_2015} and also to the density, or confining energy, of the particles \cite{filinov_2001}. Hence Wigner localization is expected for low density systems or for large 
interaction strengths.  

The Calogero and Moshinsky models are the most salient examples of 
analytically solvable models of confined particles, including the exact 
computation of the entanglement entropies 
\cite{osenda_2015,garagiola_2016,manzano_2010,dehesa_2012,benavides_2014,
koscik_2010,koscik_2013, koscik_saha_2015}, in this sense it can 
also be mentioned the spherium model \cite{toranzo_2015}, and the quasi-solvable 
Hook model \cite{taut_1993}. In particular, the Calogero model has been widely 
studied in condensed matter physics and has experienced several 
revivals \cite{Polychronakos_1993,Polychronakos_2006}, such as the discovery of 
an explicit relation of the Calogero model with the fractionary quantum hall 
effect \cite{Azuma_1994} and fractional statistics \cite{Murthy_1994}. 
{In a previous paper~\cite{garagiola_2016}, we have studied the 
behavior of the von Neumann and R\'enyi entropies of the one- and 
two-dimensional Calogero model for two particles. By considering anisotropic 
confinement in the two-dimensional case we showed that the one-dimensional 
regime is reached when the anisotropy of the trap increases, and we also 
demonstrated that the R\'enyi entropies present a non analytical behavior in 
the neighborhood of those values of the interaction strength parameter for which 
the reduced density matrix has finite support.}

Motivated by this, we consider anisotropic harmonic confinement and compute the 
exact expression for the occupation numbers or occupancies of the 
two-dimensional ground state wave function in the large interaction strength 
limit for two particles which interact via different potentials depending on the  
distance between the particles. The exact natural orbitals are obtained from the 
Schmidt decomposition of the ground state wave function in the same limit and 
the occupancies are used to evaluate several quantum information measurements 
such as von Neumann and R\'enyi entropies in closed form. {The 
method presented here is a generalization of the strategy 
developed in Refs.~\cite{koscik_2015,koscik_2010,koscik_2015_2,glasser_2013}. 
The two particle one-dimensional systems with Coulomb and inverse powers 
interactions are addressed in~\cite{koscik_2015,koscik_2015_2,glasser_2013},
while the 
natural orbitals and occupation numbers of elliptically deformed 
two-dimensional quantum dots are reported in~\cite{koscik_2010}. Here we 
give the analytical expressions of the natural 
orbitals, occupation numbers, von Neumann and R\'enyi entropies in the 
strong interaction limit for any potential which depends only on the 
interparticle distance.}

Our main purpose is to determine the influence of the anisotropy and the type of interparticle interaction 
by looking upon the linear, von Neumann, min-entropy, max-entropy 
and R\'enyi entropies as entanglement measures. We have a particular interest in which are the differences arising from a short-range interaction with respect to a long range one for which the 
emergence of Wigner Molecules has been widely described (see, for example, Refs.  
\cite{jauregui_1993,guclu_2008,cavaliere_2014}). With this aim, we study 
two interacting potential cases for each interaction range, including one that can be exactly 
solved. In the long-range interaction case we consider the inverse power and the 
inverse logarithmic potential, and for the short-range interaction we solve the 
screened inverse power potential and a Gaussian repulsive interaction. It is important to 
emphasize that the inverse power interaction case is used to model quantum dots 
\cite{kallikaos_2008} or ion traps \cite{wineland_1987} where the large 
interaction regime can be achieved experimentally due to a strong interaction 
between the particles or a weak confinement energy scale, for inverse square
power one gets the Calogero model, while the screened Coulomb interaction provides a simple model potential for ions and plasmas \cite{li_2012}.

The paper is organized as follows. The 
model is discussed in Section \ref{section_2d2p_sys}. In Section 
\ref{section_analytical_derivation} 
we show the derivation of the analytical occupancies of two interacting 
particles in a two-dimensional anisotropic harmonic trap, while in Section 
\ref{section_entropies} we calculate the entropic entanglement measures. We 
discuss the results for long- and short-range interaction potentials in Sections 
\ref{section_long_range} and \ref{section_short_range} respectively. Finally, a 
summary and conclusions are presented in Section 
\ref{section_summary_conclusions}.

\section{Confined two-dimensional two-particle systems}
\label{section_2d2p_sys}

The physics of confined particle systems is nowadays very relevant to understand the many recent experiments conducted in cold atom traps or in quantum dots, at least in a qualitative way \cite{andrei_1998, piot_2008, kallikaos_2008, ellenberger_2006, singha_2010, kristinsdottir_2011, meyer_2009}. The models for those systems contain two contributions to the potential energy: one is given by the trap potential and the other by the interaction between the particles. For small dots, containing few electrons the trap potential can be approximated by a harmonic one \cite{cavaliere_2009}, therefore we focus here on two interacting particles in a two-dimensional anisotropic harmonic traps, and implement a method to solve the entanglement spectrum in the large interaction limit for arbitrary interaction potentials. The Hamiltonian for two particles in an anisotropic trap, in atomic units, is  

\begin{equation}
\label{H_QD_2D}
H = -\frac{1}{2}\left( \nabla_1^2+\nabla_2^2\right) + \frac{1}{2}\left\lbrace (x_1^2+x_2^2)+\varepsilon^2(y_1^2+y_2^2) \right\rbrace + g V\left( r_{12} ; \left\lbrace \gamma_{i} \right\rbrace \right) \; ,
\end{equation}

\noindent where the frequency of the trap was taken equals to unity, $\varepsilon > 1$ is the anisotropy parameter, $V\left( r_{12} ; \left\lbrace \gamma_{i} \right\rbrace \right)$ denote the interaction potential as a function of the interparticle distance $r_{12}$ and some parameters $\left\lbrace \gamma_{i} 
\right\rbrace$, and $g$ is the ratio between the interaction and the confinement 
energy scale. By introducing the center of mass $\vec{R} = \frac{1}{2} (\vec{r}_1 + \vec{r}_2)=(X,Y)$ and relative coordinates $\vec{r} = \vec{r}_2-\vec{r}_1=(x,y)$ the Hamiltonian 
(\ref{H_QD_2D}) decouple as  $H = H^R + H^r$, where

\begin{eqnarray}
\label{H_cal_2D_mc}
& &H^R = -\frac{1}{4} \nabla_R^2 + \left(X^2+ \varepsilon^2 Y^2\right)\; , \\
\label{H_cal_2D_r}
& &H^r = -\nabla_r^2 + V^{eff}(x,y;\varepsilon,\left\lbrace \gamma_{i} \right\rbrace)\; , 
\end{eqnarray}

\noindent and $V^{eff}$ is the effective potential of the relative Hamiltonian given by

\begin{equation}
\label{Veff}
V^{eff}(x,y;\varepsilon,\left\lbrace \gamma_{i} \right\rbrace)= 
\frac{1}{4}\left( x^2+ \varepsilon^2 y^2\right) + g V\left( \sqrt{x^2+y^2} ; 
\left\lbrace \gamma_{i} \right\rbrace \right)\; .
\end{equation}

The total wave function is then the product of the center of mass wave function and the relative wave function 

\begin{equation}
\label{psi_prod}
\Psi(x,y,X,Y) = \psi^R(X,Y) \psi^r(x,y) \; ,
\end{equation}
 
\noindent and, consequently, the Schr\"odinger equation separates into two 
equations

\begin{eqnarray}
\label{Schr_eq_1}
& &H^R\psi^R(\vec{R}\,) = E^R \psi^R(\vec{R}\,)\; ,\\
\label{Schr_eq_2}
& &H^r\psi^r(\vec{r}\,) = E^r \psi^r(\vec{r}\,)\; .
\end{eqnarray}

The solutions of the center of mass equation (Eq. (\ref{Schr_eq_1})) are the 
eigenfunctions of the harmonic oscillator that are invariant under particle 
exchange.
%
%

The relative Hamiltonian, Eqs.~(\ref{H_cal_2D_r}) and~(\ref{Veff}), must be 
analysed on a case-specific basis. However, in 
the next section we present a method to obtain the large interaction strength 
limit of general potentials that fulfil simple requirements.

\section{Derivation of the analytical occupancies}
\label{section_analytical_derivation}

The relative wave function may be obtained by solving the Schr\"odinger 
equation in the large interaction strength regime, $g \gg 1$, by using the 
harmonic approximation (HA) \cite{james_1998, balzer_2006}. In the framework of the harmonic 
approximation one has to find the minima of the effective potential Eq. 
(\ref{Veff}) and then the potential is replaced by its Taylor 
expansion up to second order about its minima, which satisfy $\nabla V^{eff}(x,y;\varepsilon,\left\lbrace \gamma_{i} \right\rbrace) = 0 $. If the potential is repulsive, decreases monotonously 
and $ V \left( r ; \left\lbrace \gamma_{i} \right\rbrace \right) \to 
0 $ for $r\to\infty$, with $\varepsilon > 1$, the minima lie on the $x-$axis 
and can be written as

\begin{equation}
\label{min_V_eff}
\vec{r}_{min} =\left(\pm x_0, 0 \right) \;\;\;\; \mbox{with $x_0>0$ given 
by}\;\;\;\; \frac{1}{2g}=-\left.\left( \frac{1}{r} \frac{\partial V}{\partial 
r}\right)\right\vert_{x_0} \,. 
\end{equation}

\noindent It is important to notice that when the particles are confined in an 
isotropic trap, \textit{i.e} $\varepsilon=1$, the minima degenerate into a 
circle of radius $x_0$. 

Within the harmonic approximation, a Hamiltonian of uncoupled oscillators is obtained

\begin{equation}
\label{H_Ha}
H^r_{HA} = -\nabla_r^2 + \frac{1}{2} \left\lbrace \omega_x^2 \left( x- x_0  \right)^2 + \frac{1}{2} \left(\varepsilon^2 -1 \right) y^2 \right\rbrace\; , 
\end{equation}

\noindent with a frequency associated to the $x$-coordinate given by
\begin{equation}
\label{wx_2}
\omega_x^2 = \frac{1}{2} \left( 1 + \frac{\left. \frac{\partial^2 V}{\partial 
r^2}\right\vert_{x_0} }{\left. -\frac{1}{r} \frac{\partial V}{\partial 
r}\right\vert_{x_0}} \right)\;  ,
\end{equation}

\noindent where the dependence on the parameters $g$ and $\left\lbrace \gamma_{i} \right\rbrace$ is implicit in $x_0=x_0\left(g, \left\lbrace \gamma_{i} \right\rbrace \right) $. 
%
%
%
%
%

The totally symmetric ground state wave function $\Psi ^{GS} \left( \vec{r}_1, 
\vec{r}_2 \right)$ of the harmonic Hamiltonian $H^R + H^{r}_{HA}$ is a product 
of Gaussians. The eigenvalues of the one-particle reduced 
density matrix $\rho=\textrm{Tr}_2 \left(\left| \Psi^{GS}\right\rangle 
\left\langle \Psi^{GS} \right|\right)$ are explicitly obtained (see supporting 
information), and are given by

\begin{equation}
\label{auval_2D}
\Lambda_{l,\tilde{l}} = \Lambda_{l}^{x} \, \Lambda_{\tilde{l}}^{y} \;,
\end{equation}

\noindent where

\begin{equation}
\label{auval_2D_x}
\Lambda_{l}^{x}  = \frac{\left(1-\zeta\left( \omega_x\right)  \right)}{2 \left( 1 + e^{-\frac{x_0^2\,\omega_x}{\sqrt{2}}} \right) } \zeta\left( \omega_x\right)^{l} \;\;\;\; \mbox{,}\;\;\;\; \zeta\left( \omega_x\right) = \left( \frac{ \left( 2 \omega_x^2\right) ^{\frac{1}{4}} - 1}{\left( 2 \omega_x^2\right) ^{\frac{1}{4}} + 1}\right)^2 \;,
\end{equation}

\noindent and

\begin{equation}
\label{auval_2D_y}
\Lambda_{\tilde{l}}^{y} = \left(1-\xi(\varepsilon)\right) \xi(\varepsilon)^{\tilde{l}} \;\;\;\; \mbox{,}\;\;\;\; \xi(\varepsilon) = \left( \frac{\left( \varepsilon^2 -1 \right) ^\frac{1}{4} - \sqrt{\varepsilon}}{\left( \varepsilon^2 -1 \right) ^\frac{1}{4} + \sqrt{\varepsilon}}\right)^2 \;.
\end{equation}

\noindent where $l,\tilde{l}=0,1,2,\ldots$. Each eigenvalue, or occupancy, 
is doubly degenerate due to the particle exchange symmetry.

The limiting values and behavior of  $\zeta\left(\omega_x\right)$ and $\xi(\varepsilon)$ are needed to compute the entropic quantities. We note then that for $\omega_x>0$, $\zeta\left( \omega_x\right)$ 
is always below unity and $\zeta\left( \omega_x\right) \to 1$ when $\omega_x 
\to \infty$, while for $\varepsilon>1$, $\xi(\varepsilon)$ remains below one 
and $\xi(\varepsilon) \to 1$ for $\varepsilon \to 1^+$, then we must be specially careful in the isotropic confinement case (see Eq. (\ref{auval_2D_y})). For large anisotropy parameter 
$\varepsilon \gg 1 $ one gets $\xi(\varepsilon) \to 0$ and the occupancies reach 
the asymptotic values of the one dimensional model $\Lambda_{l}^{x}$.


\section{Entropies in the large interaction strength limit}
\label{section_entropies}

The entanglement can be measured using different entropic quantities. If 
$\{\Lambda_i\}$ is the complete set of eigenvalues, then the 
R\'enyi entropies are a family of such entropies defined by

\begin{equation}
\label{def_Renyi}
S^{\alpha} = \frac{1}{1-\alpha} \,\log_2 \mbox{Tr} \, \rho^{\alpha} = 
\frac{1}{1-\alpha}\, \log_2 \left( \sum_i \Lambda^{\alpha}_i \right)  \,,
\end{equation}

\noindent which are widely used in many-body or extended systems 
\cite{Calabrese2011a,Alba2010}. 
Special values of the parameter 
$\alpha$ allow to recover other entropies, being the min- and max-entropy good 
examples obtained by taking the limits $\alpha\rightarrow \infty$ and 
$\alpha\rightarrow 0$, respectively. The min-entropy serves as a lower bound to 
the entanglement measures obtained from the whole family of entropies. The 
Hartley or 
max-entropy, $S^{0}=\log_2 R$, only depends on the Schmidt rank $R$ 
of the spectrum distribution and is a measure of bipartite entanglement which 
serves as a criterion for efficient classical representation of the state 
\cite{amico_2014}. The distribution of the entanglement spectrum can be better 
understood by computing the R\'enyi entropies for many different values of the 
parameter $\alpha$ \cite{garagiola_2016}. The von Neumann entropy is given by 

\begin{equation}
\label{def_vN}
S_{vN} = - \mbox{Tr} \left(\rho \log_2 \rho \right) = - \sum_i \Lambda_i \log_2 
\Lambda_i \,.
\end{equation}

\noindent The von Neumann entropy has been used to study entanglement in 
continuous variables systems and spin models \cite{Alba2010, osenda_2007, 
pont_2010}. It can be recovered from the R\'enyi entropies in the 
limit $\alpha 
\to 1$. Finally, some authors use the linear entropy, defined by

\begin{equation}
\label{def_SL}
S_{L} = 1 - \mbox{Tr} \, \rho^2 = 1 - \sum_i \Lambda^{2}_i \,, 
\end{equation}

\noindent since for continuous variable systems the calculation of $\mbox{Tr} 
\, 
\rho^2$ is reduced to a single integral. Even though, the linear entropy has no 
relevant information for the systems studied in the present work, we compute it 
for the sake of completeness.


Once we have obtained the occupancies it is possible to calculate the quantum entropies. Since these calculations involve geometric series in 
$\zeta\left(\omega_x\right)$ and $\xi(\varepsilon)$, the limiting values must be carefully computed.

Let us start with the R\'enyi entropies defined by Eq. (\ref{def_Renyi}). It is straightforward to show that due to the separability of the wave function,  the R\'enyi entropies in the large interaction strength limit are the sum of the entropy associated to $\psi_x(x_1,x_2)$ and 
$\psi_y(y_1,y_2)$, then

\begin{equation}
\label{Renyi_xy}
S^{\alpha} = S^{\alpha}_{x}(\omega_x) + S^{\alpha}_{y}(\varepsilon) \,,
\end{equation}

\noindent where

\begin{equation}
\label{Renyi_2D_x}
S^{\alpha}_{x} \left( \omega_x\right) = \frac{1}{1-\alpha} \log_2 \left(  
\frac{(1-\zeta\left( \omega_x\right))^ \alpha}{(1-\zeta\left( \omega_x\right)^\alpha)}
\right) + 1 \,,
\end{equation}

\noindent and

\begin{equation}
\label{Renyi_2D_y}
S^{\alpha}_{y} (\varepsilon) = \frac{1}{1-\alpha} \log_2 \left(  
\frac{(1-\xi(\varepsilon))^ \alpha}{(1-\xi(\varepsilon)^\alpha)}
\right) \,.
\end{equation}


Again, due to the separability of the wave function, we can write the two-dimensional von Neumann entropy of Eq. (\ref{def_vN}) as

\begin{equation}
\label{S_VN}
S_{vN} = S_{x}^{1}\left( \omega_x\right) + S_{y}^{1}(\varepsilon) \;,
\end{equation}
\noindent where each one of the terms in the sum has the form of a 
one-dimensional von Neumann entropy \cite{koscik_2015}, {\it i.e.}

\begin{equation}
\label{Sx}
S_{x}^{1}\left( \omega_x\right) = - \frac{\log_2\left( \left(1-\zeta\left( \omega_x\right) \right)^{(1-\zeta\left( \omega_x\right))} \zeta\left( \omega_x\right)^{\zeta\left( \omega_x\right)} \right)}{ \left(1-\zeta\left( \omega_x\right)\right)} + 1\;,
\end{equation}

\begin{equation}
\label{Sy}
S_{y}^{1}(\varepsilon) = - \frac{\log_2\left( \left(1-\xi(\varepsilon)\right)^{(1-\xi(\varepsilon))} \xi(\varepsilon)^{\xi(\varepsilon)} \right)}{ \left(1-\xi(\varepsilon)\right)}\;.
\end{equation}

The super-index points that the von Neumann entropy can be obtained as a limiting case of the 
R\'enyi entropies when $\alpha\to 1$. 

It is worth to notice that from Eq. (\ref{Renyi_2D_x}) and (\ref{Renyi_2D_y}) 
it is straightforward to show that the min-entropy $S^\infty$, can also be written as a two-term sum:

\begin{equation}
\label{min_entropy}
S^\infty = \lim_{\alpha\to\infty} \left( S^{\alpha}_{x}\left( \omega_x\right) + S^{\alpha}_{y}(\varepsilon) \right)  = \lim_{\alpha\to\infty}  S^{\alpha}_{x}\left( \omega_x\right) + \lim_{\alpha\to\infty} S^{\alpha}_{y}(\varepsilon)= S^{\infty}_{x}\left( \omega_x\right)+S^{\infty}_{y}(\varepsilon) \,.
\end{equation}

The Hartley or max-entropy in the large interaction strength limit can also be calculated as a limiting case with $\alpha\to 0$, $S^{0}=\log_2 R$, and has finite value only when the one-particle reduced density matrix has finite support. 

The two-dimensional linear entropy defined by Eq. (\ref{def_SL}) gives

\begin{equation}
\label{le_final}
S_{L} = 1 - \frac{1}{2} \; \frac{1-\zeta\left( 
\omega_x\right)}{1+\zeta\left( \omega_x\right)} \; 
\frac{1-\xi(\varepsilon)}{1+\xi(\varepsilon)} \;.
\end{equation}

\noindent For the isotropic model $\varepsilon \to 1^{+}$, $\xi(\varepsilon)\to 1$ and the linear entropy goes to one, while for any other value of $\varepsilon$ the linear entropy remains below one. 

A comment on the extension of the previous results to dimension $D$ is in 
place. They can be extended if one considers $D-1$ anisotropy parameters (see 
supporting information). The von Neumann and R\'enyi entropies are the sum of 
$D$ terms each one associated to one cartesian coordinate and, as we 
demonstrated for the two-dimensional case, the $x$-entropy term depends on the 
parameters of the interaction potential through $\omega_x$ and each one of the 
remaining terms depend on only one of the $D-1$ anisotropy parameters.


\begin{figure}
\begin{center}
\includegraphics[width=0.3\textwidth]{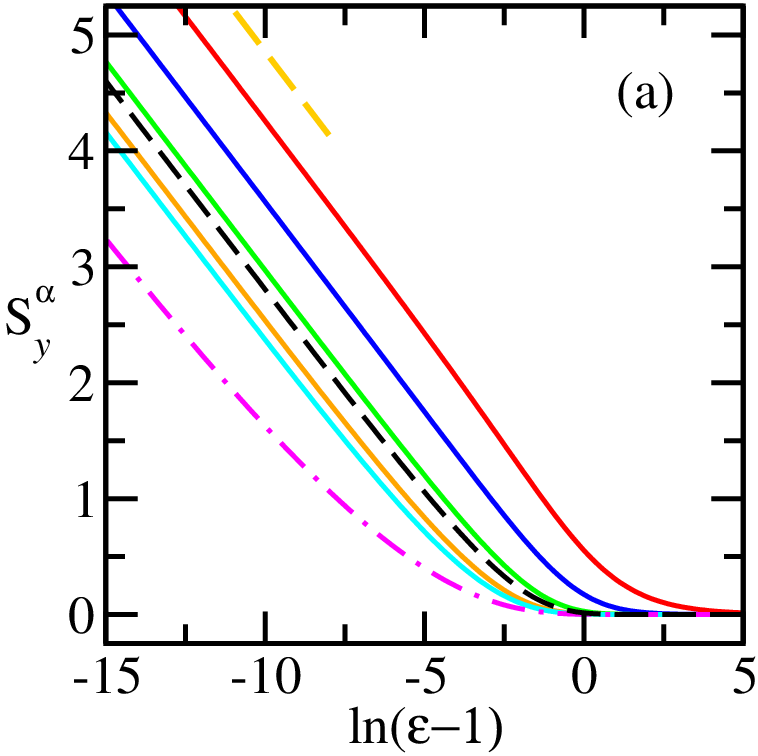}
\includegraphics[width=0.32\textwidth]{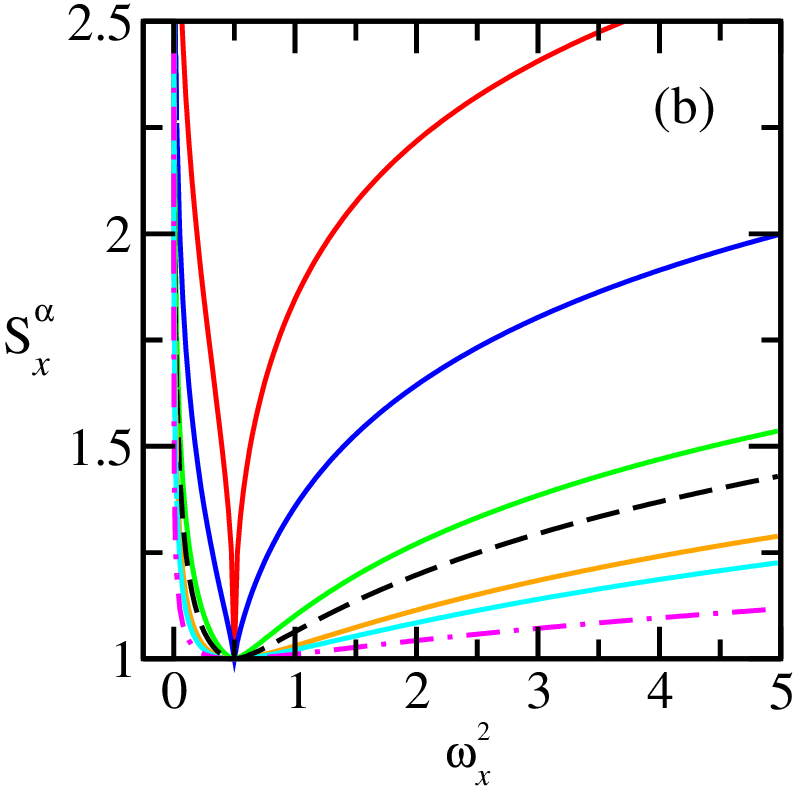}
\end{center}
\caption{\label{f_S_2D_omega}  Both terms of the two-dimensional von Neumann (black dashed), min-entropy (magenta dash-dotted) and R\'enyi entropies with $\alpha = 0.2,\,0.4,\,0.8,\,1.5,\,2$ (red, blue, green, orange and cyan full lines respectively). (a) $S_y$ as a function of the anisotropy parameter $\varepsilon$. (b) One dimensional entropies $S_x$ as a function of the squared frequency, $\omega_x^2$.
}
\end{figure}

The two terms of the two-dimensional von Neumann, min-entropy and R\'enyi entropies 
($\alpha = 0.2, \,0.4, \,0.8, \,1.5, \,2$) are depicted in Fig. \ref{f_S_2D_omega}. Let us first discuss the behavior of the entropies with respect to the anisotropy of the trap, and afterwards the influence of the interparticle interaction. 

As can be appreciated in Fig. \ref{f_S_2D_omega}(a), for the isotropic model ($\varepsilon \to 1^{+}$) the entropies $S^{\alpha}_y(\varepsilon)$ diverge logarithmically, whilst for any other values 
of $\varepsilon$ they remain finite. By calculating the first derivative of Eq. (\ref{Sy}) it is 
straightforward to show that

\begin{equation}
\label{vNe_limit}
S_{vN} \sim -\frac{\ln(\varepsilon-1)}{\ln  
16}  \;\;\;\; \mbox{for}\;\; \varepsilon \sim  1^{+}\,.
\end{equation}

\noindent This asymptotic leading term is depicted in the figure as a yellow dashed line which makes the logarithmic divergence of the isotropic von Neumann entropy evident. Actually, for $\varepsilon \to 1^{+}$ the von Neumann, min-entropy and the family of R\'enyi entropies present this same behaviour. The figure also shows that for large anisotropy parameter the entropies $S^{\alpha}_{y}(\varepsilon)$ vanish. In other words, for $\varepsilon \gg 1$ the one dimensional problem is recovered and the von Neumann, R\'enyi and min-entropy reach the one dimensional values $S_{x}^{\alpha}(\omega_x)$. 

The behavior of the $x$-entropies (denoted by $S^{\alpha}_x(\omega_x)$) as a function of the frequency is shown in Fig. \ref{f_S_2D_omega}(b). The figure shows that the entropies are decreasing functions of the frequency for $0< \omega_x^2 < 1/2$, and increasing functions for $\omega_x^2 > 1/2$. 
Actually, the entropies diverge logarithmically for large frequencies and also 
for $\omega_x \to 0$, because in these limits one gets that $\zeta(\omega_x) \to 1$.

The entropy of a given system is computed using the frequency $\omega_x$ obtained by the harmonic approximation Eq. (\ref{wx_2}). If it remains finite for large interactions parameters $g \gg 1$, then the von Neumann, min-entropy and the family of R\'enyi entropies are finite for the anisotropic model and diverge 
logarithmically for the isotropic model. In the deformed or anisotropic case the 
particles crystallize around the two classical minima of the relative 
Hamiltonian giving rise to a Wigner molecule, while for the isotropic model 
those minima degenerate into a circle, the particles are no 
longer localized around discrete minima and this lack of information is 
reflected in the divergence of the entanglement entropies. If the obtained frequency 
increases monotonously for large interactions, the von Neumann, min-entropy and 
the family of R\'enyi entropies diverge logarithmically for any anisotropy 
parameter. In this sense, the behaviour of the system is defined by the 
one-dimensional entropy $S_{x}^{\alpha}(\omega_x)$. 

The previous analysis can be understood more qualitatively by using the 
Heisenberg uncertainty principle.
The width of the Gaussian wave packet in the relative coordinate 
$\psi^{r}(\vec{r})$ (ground state of the Hamiltonian Eq. (\ref{H_Ha})) goes to 
zero 
when the frequency increases. Actually, the relative position and momentum 
uncertainty are $\Delta x^{r}_{HA} = \sqrt{\langle \left( x_2 - x_1 \right)^2  
\rangle - \langle x_2 - x_1 \rangle^2} = 2^{\frac{1}{4}}/\sqrt{\omega_x}$ and 
$\Delta p^{r}_{HA} = \sqrt{\omega_x}/2^{\frac{5}{4}}$, then if $\omega_{x} \to 
0$ we obtain that $\Delta x^{r}_{HA} \to \infty$ and, conversely, when $\omega_{x} \to \infty$ 
it is straightforward to show that $\Delta p^{r}_{HA} \to \infty$.
Thus, we see that for $\omega_{x} \to \infty$ the entropy of the Wigner molecule diverge 
because the position is completely determined and consequently the momentum 
uncertainty diverges, we refer to this limit as \emph{strong 
crystallization}. The opposite case, $\omega_{x} \to 0$ leads to a 
well defined momentum state and hence we have no knowledge of the position. In 
both cases the divergence in the position or momentum width leads to the 
divergence of the entanglement entropies. Furthermore, the divergence of the 
$y$-entropies could also be explained in a similar way: for the isotropic model 
the minima degenerate into a circle and consequently the particles are no longer 
localized around any definite angular positions, but the state has definite 
angular momentum. 

\begin{figure}
\begin{center}
\includegraphics[height=0.29\textwidth]{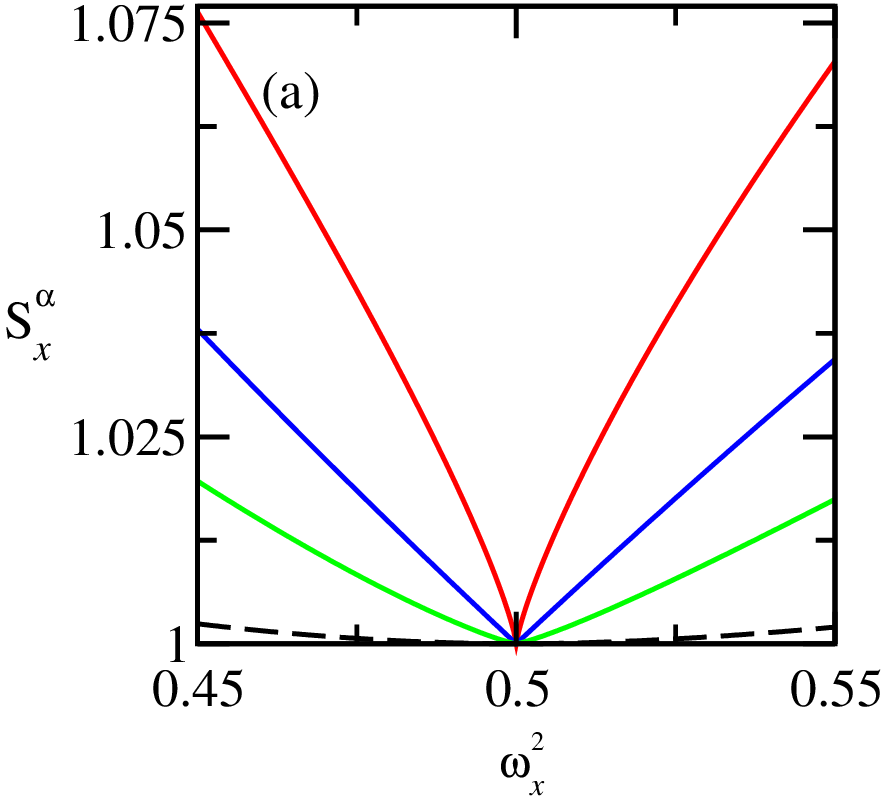}
\includegraphics[height=0.30\textwidth]{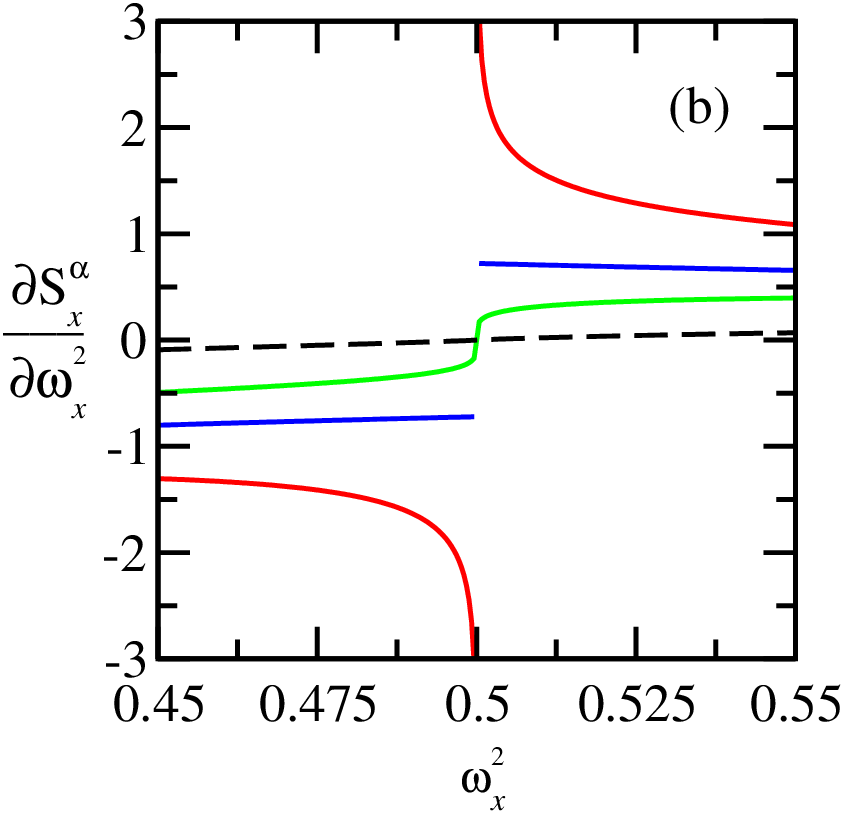}
\end{center}
\caption{\label{f_S_x_der} (a) Entropy terms $S^{\alpha}_x(\omega_x)$ and (b) their derivatives around the point $\omega_x^2=1/2$.  The von Neumann (black dashed), and R\'enyi entropies with $\alpha = 0.4,\,0.5,\,0.6$ (red, blue and green full lines respectively) are shown.
}
\end{figure}

For $\omega_x^2=1/2$ the entropies have 
their minimum value equal to unity. Around this point, the von Neumann, 
min-entropy and R\'enyi entropies with $\alpha>1$ present an analytical 
behaviour while the R\'enyi entropies with $\alpha<1$ have a non-analytical 
behaviour. The von Neumann and R\'enyi entropies with $\alpha = 0.4,\,0.5,\,0.6$ 
and their first derivatives around the point $\omega_x^2=1/2$ are shown in Fig. 
\ref{f_S_x_der} (a) and (b) respectively. It shows that the R\'enyi entropies present an infinite 
derivative for $\alpha=0.4$, discontinuous derivative for $\alpha=0.5$ and a 
continuous derivative with infinite second derivative for $\alpha=0.6$, while 
the von Neumann entropy ($\alpha\rightarrow {}^{+}1$) is an 
analytical function of the frequency.

Recent studies by Amico and co-workers in $1/2$-spin chains show the physical 
implications of non-monotonous properties of the R\'enyi entropies in many-body 
systems with topological order due to a truncation of the support of the reduced 
density matrix \cite{amico_2013,amico_2013_2,amico_2014_2}. In 
Ref.~\cite{osenda_2015} some of the present authors found that the Calogero 
model in one dimension has a finite number of non-zero occupancies for a 
discrete set of values of the interaction parameter, and in Ref. 
\cite{garagiola_2016} we demonstrated that in those particular values of 
the interaction parameter the R\'enyi entropies present a non-analytical 
behaviour. 

Summarizing, non-analytical behaviour of the R\'enyi entropies exposes the 
finite support of the reduced density matrix. In the present case, taking 
$\omega_x^2=1/2$ in Eq. (\ref{auval_2D_x}) it is straightforward 
to see that for this particular frequency there is only two non vanishing 
occupancies $\Lambda^{x}_{0}$ associated to the two lowest natural orbitals in 
the $x$-coordinate. 

In the following sections we apply our findings to study the behavior of the occupancies and entropic entanglement measures in the large interaction strength limit for different cases divided as long- or short-range potentials. From now on we calculate only the one dimensional entropy $S_{x}^{\alpha}(\omega_x)$, since the behaviour of the entropy terms $S^{\alpha}_{y}(\varepsilon)$ were already 
analysed.

\section{Long-range interaction potentials}
\label{section_long_range}

In the present section we consider two long-range interactions to exemplify our results: the inverse power interaction and inverse logarithmic interaction. 

\subsection{Inverse power interaction}

The inverse power potential is 

\begin{equation}
\label{V_inv_power}
V^{ip}\left( r ; \beta \right) = \frac{1}{r^{2\beta}}\; .
\end{equation}

\noindent For this potential $x_0$ and $\omega_x$, Eqs. (\ref{min_V_eff}) and (\ref{wx_2}), can be obtained exactly and give

\begin{equation}
\label{min_freq_V_inv_power}
x_0=(4g\beta)^{\frac{1}{2(\beta+1)}} \;\;\;\; \mbox{and}\;\;\;\;  \omega_x^2=\beta + 1\,. 
\end{equation}

\noindent Thus, $x_0$ increases when increasing the interaction strength 
parameter $g$, but the frequency remains invariant. For $\beta = \frac{1}{2},1$ one gets the Hook and the Calogero model respectively.

\begin{figure}
\begin{center}
\includegraphics[height=0.3\textwidth]{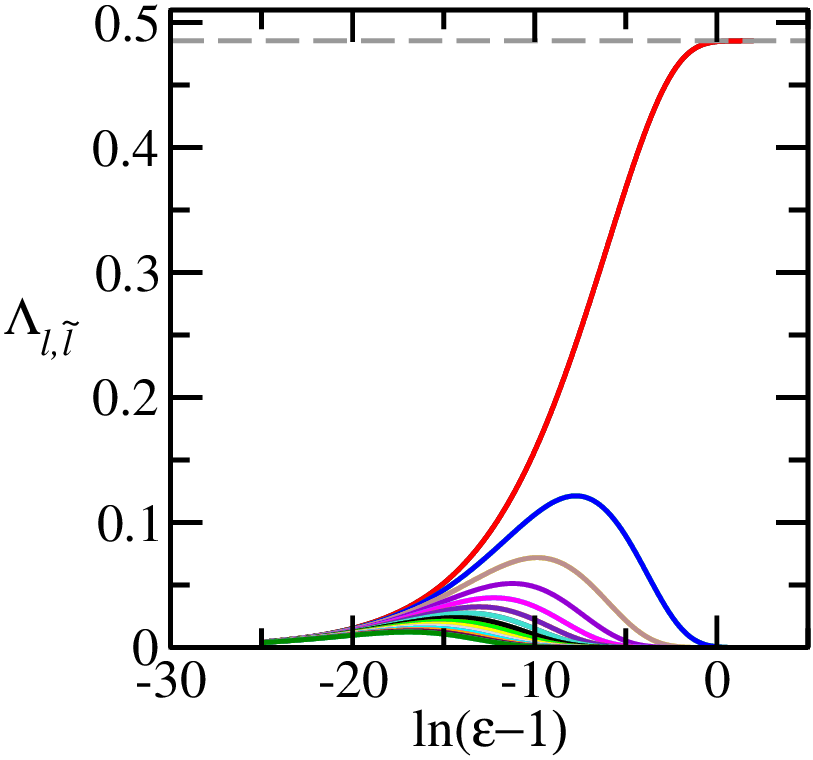}
\end{center}
\caption{  \label{f_auval_2D} Occupancies as a function of $\ln 
(\varepsilon-1)$ (see Eq. (\ref{auval_2D})) obtained for large interaction 
strength parameter, $g \gg 1$. $\Lambda_{l,\tilde{l}}$ with $l=0$ and 
$\tilde{l}=0,1,2,...,20$ from top to bottom. The dominant one-dimensional 
eigenvalue is also shown (grey dashed line) \cite{koscik_2015}.}
\end{figure}

Let us start with the Calogero model. The occupancies 
defined in Eq. (\ref{auval_2D}) for large interaction strength parameter 
$g\gg1$ are shown as a function 
of $\ln (\varepsilon-1)$ in Fig. \ref{f_auval_2D}, where the grey dashed line 
is the dominant one-dimensional occupancy in the large interaction strength 
limit, obtained from Eq. (\ref{auval_2D_x}), that is in agreement with the 
value reported in Ref. \cite{koscik_2015}. The figure shows that for $\epsilon 
\to 1^{+}$ (isotropic model) the occupancies go to zero, but note that 
their sum is always equal to $1/2$ due to the mentioned 
double degeneracy \cite{koscik_2010}. When the 
anisotropy increases all the occupancies $\Lambda_{l,\tilde{l}}$ with 
$\tilde{l}\neq 0$ present a local maximum. For fixed $l$ the value of the 
anisotropy parameter at which the maximum occurs decreases when $\tilde{l}$ 
increases, while for fixed $\tilde{l}$ this value is the same for each 
$l$. For $\varepsilon \gg 1$ the 
Hamiltonian reduces to a one dimensional oscillator and the occupancies 
$\Lambda_{l,0}$ reach the asymptotic values of the one dimensional model. For 
values of $\varepsilon$ near $\varepsilon_c = \sqrt{5}$ the occupancies with 
$\tilde{l}=0$ stabilize on the one dimensional values and those with 
$\tilde{l}\neq 0$ saturate at vanishingly small values. This feature can be 
explained if one takes into account that for $\varepsilon =\varepsilon_c$ the 
relative Hamiltonian Eq. (\ref{H_Ha}) reduces to a harmonic oscillator in polar 
coordinates around each minimum. More 
generally, for arbitrary $\beta$, the one dimensional regime is reached at the value $\varepsilon_c = \sqrt{2 \left(\beta+1\right)+1} $. In this case the effective potential of the relative 
Hamiltonian Eq. (\ref{Veff}) is isotropic in a small neighborhood around its minima. 
For $\varepsilon > \varepsilon_c$, the largest occupancy $\Lambda_{00}$ reaches 
the value $\sim 0.4853$, and the sum of 
all the remaining occupancies is only $\sim 0.0147$; this means that the two natural orbitals associated to this eigenvalue are the only two that are occupied while all the others natural orbitals contribution are negligible, and consequently, the spatial wave functions are quite similar to those two natural orbitals \cite{koscik_2010}.

\begin{figure}
\begin{center}
\includegraphics[height=0.3\textwidth]{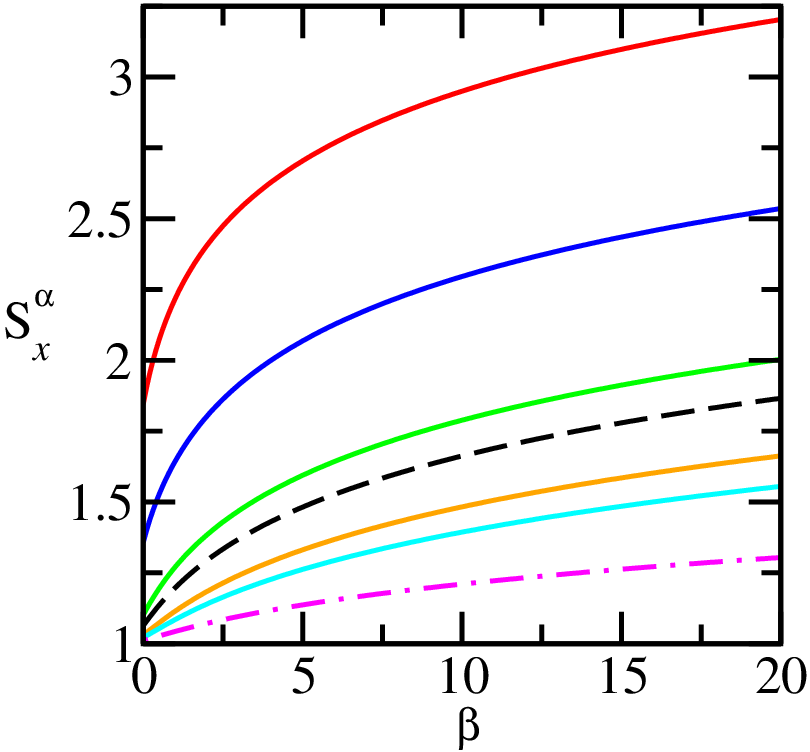}
\end{center}
\caption{\label{f_Ss} One-dimensional entropy terms $S^{\alpha}_x$ obtained for large interaction strength $g \gg 1$, as a function of the exponent of the interaction between particles, $\beta$. The von Neumann (black dashed), min-entropy (magenta dash-dotted) and R\'enyi entropies with $\alpha = 0.2,\,0.4,\,0.8,\,1.5,\,2$ (red, blue, green, orange and cyan full lines respectively) are shown.
}
\end{figure}

As we mentioned above, the dependence with $\beta$ is present only 
through the entropy term $S^{\alpha}_x(\omega_x)$. The width of the Gaussian wave packet in the relative coordinate is finite, $\Delta x^{r}_{HA} = 2^{\frac{1}{4}}/(\beta + 1)^\frac{1}{4}$ and consequently the von Neumann, min-entropy and R\'enyi entropies are finite. However, the max-entropy diverges due to the infinite support of the one-particle density matrix. Notice that in the limit $\beta \to \infty$ the entropies diverge logarithmically 
due to the divergence in the momentum uncertainty. This behavior can be seen in Fig. \ref{f_Ss} where the von Neumann, the min-entropy and R\'enyi entropies are depicted as a function of the parameter $\beta$. The R\'enyi entropies increases for decreasing $\alpha$, and the von Neumann entropy is a limiting case with $\alpha \to 1$. It is important to emphasize that taking the limit $\beta \to 0$ in the entropies does not result in the same entropies obtained for a system with harmonic confinement and a constant interaction (Eq. (\ref{V_inv_power}) with $\beta=0$), since this limit does not commute with the large interaction limit. 


\subsection{Inverse logarithmic interaction}

The potential for inverse logarithmic interparticle interaction is

\begin{equation}
\label{V_inv_log}
V^{il}\left( r \right) = \frac{1}{\ln (r+1)}\; .
\end{equation}

\noindent In this case $x_0$ and $\omega_x$ satisfy the following equations

\begin{equation}
\label{min_freq_V_inv_log}
2g = x_0 \left( x_0+1 \right) \ln^{2} \left( x_0+1 \right)  \;\;\;\; \mbox{and}\;\;\;\;  \omega_x^2= \frac{1}{2} \left\lbrace 1 + \left( \frac{\frac{2}{\ln \left( x_0+1 \right)}+1}{\frac{1}{x_0}+1} \right) \right\rbrace \,. 
\end{equation}

\noindent For large interaction strength parameter the value of $x_0$ increases 
when $g$ increases, and consequently, the frequency goes to unity. Therefore, 
for $g\gg 1$, the one-dimensional von Neumann 
and R\'enyi entropies with $\alpha>0$ remain finite, but once more the 
max-entropy diverges. We included two figures in the supporting 
information showing the qualitative behavior 
of $x_0$ and the R\'enyi entropy as a function of $\alpha$. The same 
analysis performed in the inverse power interaction case can be done for the 
inverse logarithmic potential.



\section{Short-range interaction potentials}
\label{section_short_range}

In the present section we consider two particles in a two-dimensional 
anisotropic harmonic trap with two different short-range interactions: the screened inverse 
power interaction and Gaussian repulsive interaction.

\subsection{The screened inverse power interaction}

For the screened inverse power interaction the potential is

\begin{equation}
\label{V_screened_inv_power}
V^{sip}\left( r ; \lbrace\beta,\gamma\rbrace \right) = \frac{e^{-\gamma r}}{r^{2\beta}}\; ,
\end{equation}

\noindent where $1/\gamma$ is the cut-off distance. In this case $x_0$ and $\omega_x$ are given by

\begin{equation}
\label{min_freq_V_screened_inv_power}
2g= \frac{e^{\gamma x_0} x_0^{2(1+\beta)}}{2\beta+\gamma x_0}  \;\;\;\; \mbox{and}\;\;\;\;  \omega_x^2= \frac{1}{2} \left( 1+ \frac{2 \beta}{2 \beta +\gamma x_0} + 2\beta +\gamma x_0\right)  \,. 
\end{equation}

\begin{figure}
\begin{center}
\includegraphics[height=0.3\textwidth]{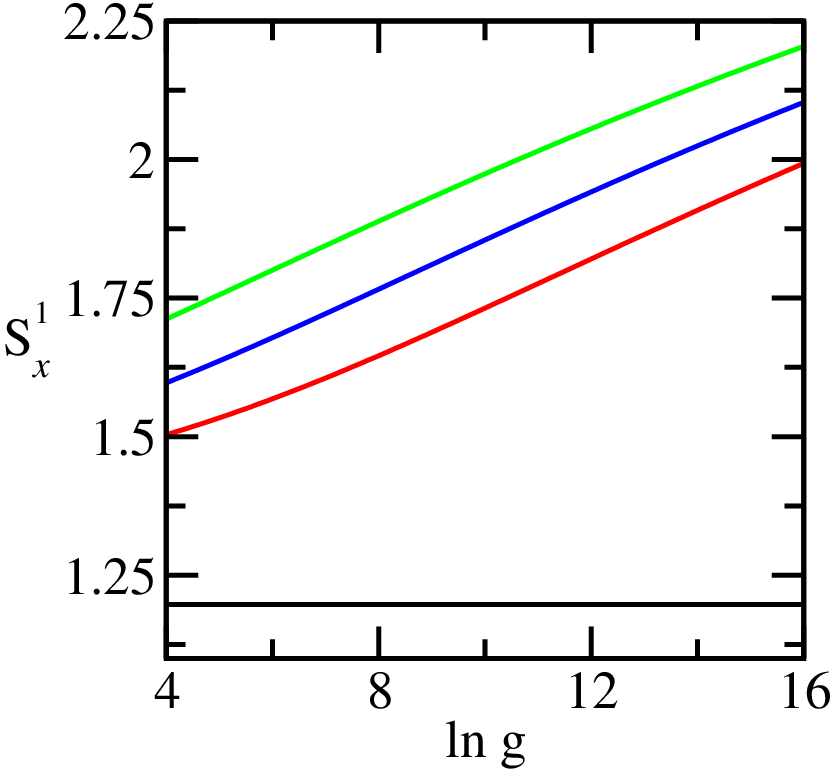}
\end{center}
\caption{  \label{f_S_screened_inverse_power} One-dimensional von Neumann entropy $S_x^{1}$ as a function of the interaction strength parameter $g$, with $\beta=1$ and $\gamma = 0,\, 1/2,\,1,\;2$ (from bottom to top, black, red, blue and green lines).
}
\end{figure}

\noindent Notice that taking $\gamma=0$ the minima and the frequency of the 
inverse power interaction is recovered, and for $\beta=0$ the interaction has 
exponential decay. For large interaction strength parameter $g$ the minima 
and the frequency increase monotonously with $g$, and consequently, the one-dimensional von Neumann and R\'enyi entropies diverge logarithmically. As we mentioned in section \ref{section_entropies} the divergence of the entropies can be explained as arising from the momentum uncertainty $\Delta p^{r}_{HA} = \sqrt{\omega_x}/2^{\frac{5}{4}}$, which diverges when $\omega_{x} \to \infty$. Actually, the larger the $\gamma$ parameter is, the larger the frequency is and the higher the entanglement 
entropies are, this behaviour is shown in Fig. 
\ref{f_S_screened_inverse_power} where the one-dimensional von Neumann entropy 
is depicted as a function of the interaction strength 
for $\beta=1$ and $\gamma = 0,\, 1/2,\,1,\;2$. 

\subsection{The Gaussian repulsive interaction}

In this subsection we consider the following interaction potential 

\begin{equation}
\label{V_Gaussian}
V^{gr}\left( r ; \sigma\right) = e^{-\frac{r^2}{2\sigma^2}} \; ,
\end{equation}

\noindent where $\sigma$ is the half width of the potential. In this case $x_0$ and $\omega_x$ can be found exactly

\begin{equation}
\label{min_freq_V_Gaussian}
x_0 = \sigma \sqrt{2 \ln \left( \frac{2 g}{\sigma^2} \right)}   \;\;\;\; \mbox{and}\;\;\;\;  \omega_x^2= \frac{1}{2} \frac{x_0^2}{\sigma^2} = \sqrt{ \ln \left( \frac{2 g}{\sigma^2} \right)} \;\;\;\; \mbox{with}\;\;\;\; g \geq \frac{\sigma^2}{2}\,. 
\end{equation}

\noindent They are increasing functions of the interaction strength parameter $g$, thus, the one-dimensional von Neumann and R\'enyi entropies in the large interaction strength limit diverge 
logarithmically. We interpret this divergence in the same way as for the 
screened inverse power interaction. It is worth to mention that the limit 
$\sigma \to 0$ does not reproduce the results of a delta interaction, which have a finite von Neumann entropy \cite{avakian_1987}, since this limit does not commute with the large interaction strength limit. 

From Eq. (\ref{min_freq_V_Gaussian}) it is straightforward to show that for $g=g_c$ with 

\begin{equation}
\label{g_c_Gaussian}
g_c = \frac{ \sigma^2 e^{\frac{1}{2}}}{2}\,, 
\end{equation}

\begin{figure}
\begin{center}
\includegraphics[height=0.3\textwidth]{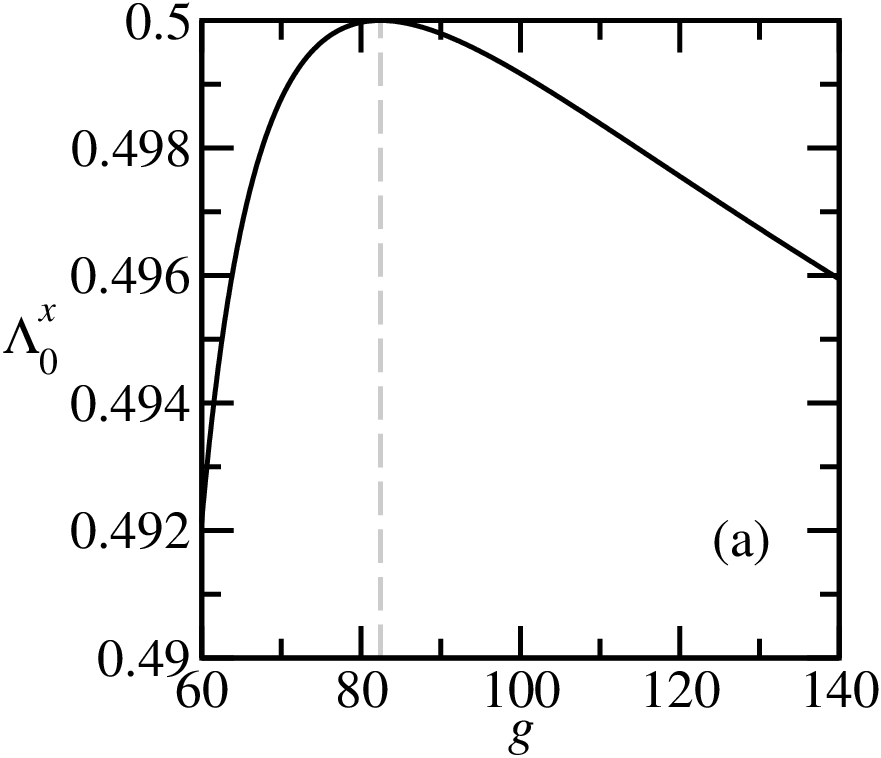}
\includegraphics[height=0.3\textwidth]{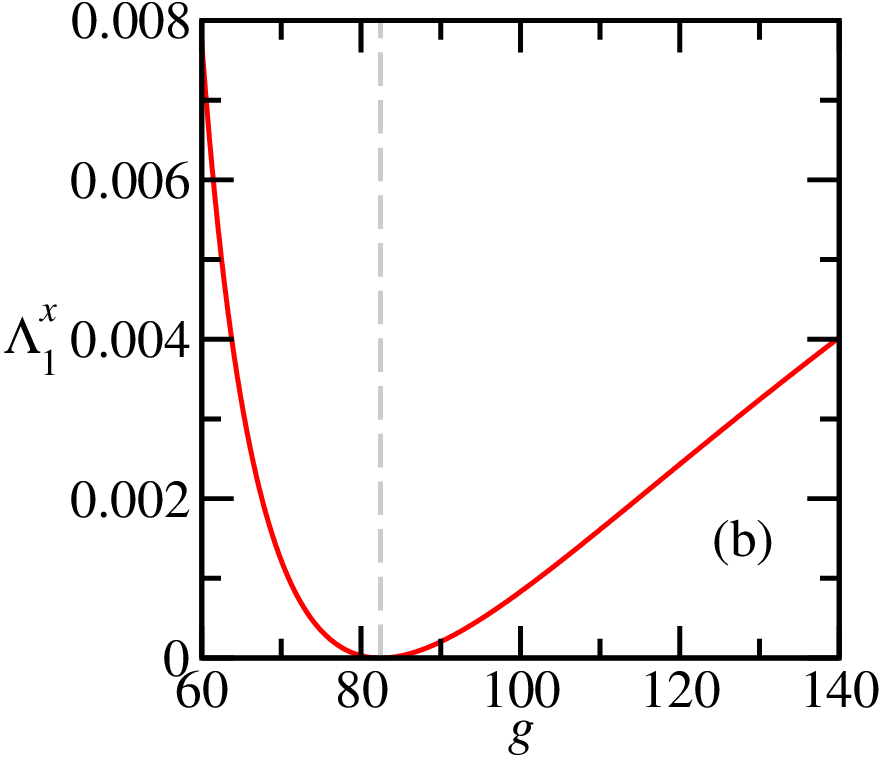}
\end{center}
\caption{  \label{f_occupancies_Gaussian} One-dimensional occupancies $\Lambda_{l}^{x}$ (Eq. \ref{auval_2D_x}) with (a) $l=0$ and (b) $l=1$, for $\sigma=10$, the value $g_c$ is depicted as a gray dashed line.
}
\end{figure}

\noindent $x_0 = \sigma$ and $\omega_x^2= \frac{1}{2}$, therefore, as was 
explained in section \ref{section_analytical_derivation} all 
the occupancies vanish except two of them. The reduced 
density matrix has then a finite support and the R\'enyi entropies 
with $\alpha <1$ have a non-analytical behaviour, while the von Neumann entropy and R\'enyi entropies with $\alpha>1$ present a minimum at $g=g_c$. The behaviour of the first and second degenerate
occupancies (Eq. (\ref{auval_2D_x})) for $\sigma=10$ is depicted in Fig. 
\ref{f_occupancies_Gaussian} together with the value $g_c$ as a gray dashed 
line. The first two occupancies $\Lambda_0^x$ (Fig. 
\ref{f_occupancies_Gaussian}(a)) reach the maximum value $1/2$ for $g=g_c$, 
value at which all the others occupancies vanish as can be appreciated for 
$\Lambda_1^x$ in Fig. \ref{f_occupancies_Gaussian}(b). 

Summarizing, we found that two trapped particles with a Gaussian repulsive 
interaction between them, have a reduced density matrix with infinite support 
(infinite non-vanishing occupancies) for all the interaction
strengths but for $g=g_c$, value at which all the occupancies vanish except 
two, and the support is finite. Nevertheless, it is important to notice that the 
largest occupancy $\Lambda_0^x \gtrsim 0.49$ throughout the considered range of 
$g$ values, and the sum of all the remaining occupancies is $\lesssim 0.01$. 
Therefore, in the neighborhood of $g_c$, the two natural orbitals associated to 
$\Lambda_0^x$ are the main contributions to the spatial wave function expansion.

\section{Summary and Conclusions}
\label{section_summary_conclusions}

In this work we present analytical expressions in the large interaction strength limit for the occupancies and quantum entropies for the ground state of a two-particle Wigner molecule in a two-dimensional anisotropic harmonic trap. Our main result is that one is able to determine the influence of the anisotropy 
and the range of the interparticle interaction looking upon the entropic entanglement measures. 

The wave function is obtained within the framework of the harmonic approximation 
for large interaction strength values, and once we have the ground state 
wave function, we calculate the occupancies from the Schmidt decomposition of the 
reduced density matrix. We obtain doubly degenerate occupancies, and relate this to 
the equivalence between particle exchange in the wave function and the exchange 
between the two minima of the effective potential of the relative Hamiltonian. 
The linear, von Neumann, min-entropy, max-entropy and R\'enyi entropies are 
calculated exactly in terms of the occupancies as a function of the anisotropy parameter and the parameters of the interaction potential. 

We found that, due to the coordinate separability of the wave function, the von Neumann, min-entropy, max-entropy and R\'enyi entropies are a sum of terms associated to each coordinate, and that only one of these 
terms depends on the anisotropy parameter and the other term is associated to 
the interaction potential. Consequently, the behavior of the entropies with 
respect to the anisotropy parameter can be analyzed without regard of the 
interaction potential, and the dependence on the interaction potential is 
entirely defined by the frequency obtained by the harmonic approximation of the 
one-dimensional problem. Moreover, we generalize these results to 
dimensions higher than two, see details in the supporting information.

We show that when the frequency remains finite for large 
interactions, then the von Neumann, min-entropy and the family of 
R\'enyi entropies remain finite for the anisotropic model and diverge 
logarithmically for the isotropic model. The divergence of the 
entanglement measure entropies of the isotropic model can be understood as follows: in the 
deformed or anisotropic case the particles locate around the two classical 
minima of the relative Hamiltonian forming a Wigner molecule, while for the 
isotropic model those minima degenerate into a circle, the 
particles are no longer localized and this lack of 
information is reflected in the divergence of the entanglement entropies. If the 
frequency increases monotonously for large interactions then, the von Neumann, 
min-entropy and the family of R\'enyi entropies diverge logarithmically for any 
anisotropy parameter. In this sense, the influence of the interaction potential 
is present only in the one-dimensional entropies.

The previous interaction-independent analysis allows us to apply 
them to different interactions straightforwardly. We group the interactions 
into short and long-range potentials and show the differences of the 
obtained results between the groups. For long-range interaction potentials, the frequencies remain finite in the large interaction strength limit, and the von Neumann, min-entropy and R\'enyi entropies are finite. In contradistinction, for short-range interaction potentials, the frequencies increase monotonously as a function of the interaction strength and, consequently, the one-dimensional von Neumann and R\'enyi entropies diverge in the large interaction strength limit. The divergence of the entanglement entropies can be explained as arising from the momentum uncertainty divergence at large frequencies. It is important to mention that the one-dimensional von Neumann, min-entropy and R\'enyi entropies of the inverse power interaction model diverge logarithmically when the power of the inverse interaction increases (see Eqs. (\ref{Renyi_2D_x}) and (\ref{Sx})), since in this limit the interaction between the particles goes to a short range one.

We also demonstrate that when the frequency associated to the interaction potential, satisfy $\omega_x^2=1/2$, the entropies have their minimum value equal to unity. Actually, the von Neumann, min-entropy and R\'enyi entropies with $\alpha>1$ present an analytical behaviour around this point, while the R\'enyi entropies with $\alpha<1$ have a non-analytical behaviour, which exposes the finite support of the reduced density matrix. For this particular frequency there is only two non vanishing occupancies $\Lambda^{x}_{0}$. Similar features were also recently reported by Amico and co-workers for $1/2$-spin chains \cite{amico_2013,amico_2013_2,amico_2014_2}, and for the Calogero model \cite{garagiola_2016} by the present authors. We further illustrate these features showing that two trapped particles with a Gaussian repulsive interaction have a reduced density matrix with infinite support (infinite non-vanishing occupancies) for all Hamiltonian parameters except for those values that allow $\omega_x^2=1/2$, where all the occupancies vanish except two, and the reduced density matrix has finite support.

As a final remark, there is a very recent work concerning a 
system of two Coulombically interacting particles confined to a $D-1$ sphere, 
where the dependence of the entanglement measures on the radius of the system 
and the spatial dimensionality has been investigated \cite{toranzo_2015}. Thus, 
as future perspectives we would like to study the effects of the dimensionality 
and the interaction strength on the entanglement of two confined particles which 
interact via a general potential, taking as a starting point the results 
obtained in the generalization to dimensions higher than two presented in the 
second section of the supporting information.

\section*{Acknowledgments}
We acknowledge SECYT-UNC and CONICET for partial financial support of this 
project. E.C. would like to thank Alvaro Cuestas for an exhaustive reading of 
the manuscript.


\newpage

\noindent {\bf \large Supporting Information\\ Long- and short-range interaction footprints in 
entanglement entropies of 
two-particle Wigner molecules in 2D quantum traps}

\addtocounter{section}{-7}

\section{Derivation of the analytical occupancies and natural orbitals}
\label{sup_analytical_derivation}

We obtain the relative wave function by solving the Schr\"odinger equation 
in the large interaction strength regime, within the harmonic approximation (HA) \cite{sjames_1998, 
sbalzer_2006}. As is pointed in the introduction, the method 
presented here is a generalization of the strategy developed in Ref. 
\cite{skoscik_2015,skoscik_2010,skoscik_2015_2,sglasser_2013} for some particular 
interaction potentials (Coulomb and inverse powers) to any interaction which 
depends only on the interparticle interaction. As is also 
mentioned in the main text, if the 
potential is repulsive, decreases monotonously 
and $ V \left( r ; \left\lbrace \gamma_{i} \right\rbrace \right) \to 
0 $ for $r\to\infty$, with $\varepsilon > 1$, the minima lie on the $x-$axis 
and can be written as

\addtocounter{equation}{-34}

\begin{equation}
\label{smin_V_eff}
\vec{r}_{min} =\left(\pm x_0, 0 \right) \;\;\;\; \mbox{with $x_0>0$ given 
by}\;\;\;\; \frac{1}{2g}=-\left.\left( \frac{1}{r} \frac{\partial V}{\partial 
r}\right)\right\vert_{x_0} \,. 
\end{equation}

Within the harmonic approximation, a Hamiltonian of uncoupled oscillators is 
obtained

\begin{equation}
\label{sH_Ha}
H^r_{HA} = -\nabla_r^2 + \frac{1}{2} \left\lbrace \omega_x^2 \left( x- x_0  
\right)^2 + \frac{1}{2} \left(\varepsilon^2 -1 \right) y^2 \right\rbrace\; , 
\end{equation}

\noindent with a frequency associated to the $x$-coordinate given by

\begin{equation}
\label{wx}
\omega_x^2 = \frac{1}{2} + g \left.\left(  \frac{\partial^2 V}{\partial 
r^2}\right)\right\vert_{x_0}\; . 
\end{equation}

\noindent By using Eq. (\ref{smin_V_eff}), the frequency can be rewritten as

\begin{equation}
\label{swx_2}
\omega_x^2 = \frac{1}{2} \left( 1 + \frac{\left. \frac{\partial^2 V}{\partial 
r^2}\right\vert_{x_0} }{\left. -\frac{1}{r} \frac{\partial V}{\partial 
r}\right\vert_{x_0}} \right)\; . 
\end{equation}

\noindent where the dependence on the parameters $g$ and $\left\lbrace 
\gamma_{i} \right\rbrace$ is implicit in $x_0=x_0\left(g, \left\lbrace 
\gamma_{i} \right\rbrace \right) $. 

The solutions to the corresponding Schr\"odinger equation are

\begin{equation}
\label{sol_rel}
\psi_{\tilde{n},\tilde{m}}^r(\vec{r}) = 
e^{-\frac{\omega_x}{\sqrt{2}}\frac{(x-x_0)^2}{2}}H_{\tilde{n}}\left(\sqrt{\frac{
\omega_x}{\sqrt{2}}}(x-x_0) \right) 
e^{-\frac{\sqrt{\varepsilon^2-1}}{4}y^2}H_{\tilde{m}}\left( 
\left(\frac{\varepsilon^2-1}{4} \right)^{1/4} y\right)\; ,
\end{equation}

\noindent with energies

\begin{equation}
\label{energy_rel}
E_{\tilde{n},\tilde{m}}^{r} = \sqrt{2} \omega_x \left(\tilde{n} + \frac{1}{2} 
\right) + \sqrt{\varepsilon^2 -1} \left(\tilde{m} + \frac{1}{2} \right)\; .
\end{equation}

By using the solutions of the center of mass equation, given by

\begin{equation}
\label{sol_CM}
\psi_{n,m}^R(\vec{R}) = e^{-X^2}H_n\left( \sqrt{2}X \right) e^{-\varepsilon 
Y^2}H_m\left( \sqrt{2\varepsilon}Y\right)\; ,
\end{equation}

\noindent with energies 

\begin{equation}
\label{sol_CM_energies}
E_{n,m}^R = \left(n+\frac{1}{2}\right)+\varepsilon \left(m+\frac{1}{2}\right) \; 
,
\end{equation}
\noindent and (\ref{sol_rel}) taking $n=m=\tilde{n}=\tilde{m}=0$, the totally 
symmetric ground state wave function is 

\begin{equation}
\label{total_wf}
\Psi ^{GS} \left( \vec{r}_1, \vec{r}_2 \right) = C \,  e^{-\varepsilon 
\frac{(y_1+y_2)^2}{4}}           e^{-\frac{\sqrt{\varepsilon ^2 -1} 
(y_2-y_1)^2}{4}} 
e^{-\frac{\left(x_1+x_2\right)^2}{4}}        \lbrace    
e^{-\frac{\omega_x}{\sqrt{2}} \frac{\left(x_2-x_1-x_0\right)^2}{2}} +  
e^{-\frac{\omega_x}{\sqrt{2}} \frac{\left(x_2-x_1+x_0\right)^2}{2}}    \rbrace 
\;,
\end{equation}

\noindent where $C$ is the normalization constant

\begin{equation}
\label{normalization}
C = \left(
\frac{\sqrt{\omega_x }}
{2^{\frac{3}{4}} \pi 
\left( 1+e^{-\frac{x_0^2 \, \omega_x}{\sqrt{2}}}\right)} 
\right)^{\frac{1}{2}} \, 
\left(
\frac{\sqrt{\varepsilon \sqrt{\varepsilon^2-1}}}
{\pi}
\right)^{\frac{1}{2}}
\; .
\end{equation}

The total wave function, Eq. (\ref{total_wf}), is separable in the $x$ and $y$ 
coordinates as $\Psi^{GS} \left( \vec{r}_1, \vec{r}_2 \right) = \psi_x(x_1,x_2) 
\psi_y(y_1,y_2)$ where,

\begin{equation}
\label{psi_x}
\psi_x(x_1,x_2) = C_x \; \left\lbrace 
q\left(x_1-\frac{x_0}{2},x_2+\frac{x_0}{2}\right)+q\left(x_1+\frac{x_0}{2},
x_2-\frac{x_0}{2}\right)\right\rbrace \; ,
\end{equation}

\noindent with 

\begin{equation}
\label{function_q}
q(u,v) = e^{-\frac{1}{4}\left(1+\sqrt{2} \omega_x \right) 
\left(u^2+v^2\right) - \frac{1}{2}\left(1-\sqrt{2} \omega_x \right) u v} \; ,
\end{equation}

\noindent and

\begin{equation}
\label{psi_y}
\psi_y(y_1,y_2) = C_{y}\, 
e^{-\frac{\varepsilon+\sqrt{\varepsilon^2-1}}{4}\left(y_1^2+y_2^2\right) - 
\frac{\varepsilon-\sqrt{\varepsilon^2-1}}{2}y_1y_2} \; ,
\end{equation}

\noindent where $C_{x}$ and $C_{y}$ are the first and second factors of Eq. 
(\ref{normalization}) respectively. 

Since we are interested in the occupancies of the 
natural orbitals, we solve the integral equation for the eigenvalues 
of the one-particle reduced density matrix obtained 
from the totally symmetric ground state $\Psi^{GS} 
\left( \vec{r}_1, \vec{r}_2 \right)$, Eq. (\ref{total_wf}). 
\begin{equation}
\label{int_eq_rho}
\int \rho \left( \vec{r}_1, \vec{r}\,'_1 \right) \phi_i\left(\vec{r}\,'_1 
\right) d\vec{r}\,'_1 = \Lambda_i \phi_i\left(\vec{r}_1 \right) \;,
\end{equation}
It is possible to reduce the computation effort by noting that the iterated 
kernel ($\rho$) of a symmetric 
kernel ($\Psi^{GS}$) has the same eigenfunctions as the kernel, while the 
iterated eigenvalues are the squared eigenvalues of the kernel 
\cite{stricomi_1957}. This means that instead of directly solving 
Eq. (\ref{int_eq_rho}) one can solve the 
following equation    

\begin{equation}
\label{int_eq_psi}
\int \Psi^{GS} \left( \vec{r}_1, \vec{r}_2 \right) \phi_i\left(\vec{r}_2 
\right) 
d\vec{r}_2 = \lambda_i \phi_i\left(\vec{r}_1 \right) \;,
\end{equation}

\noindent with $\Lambda_i = \lambda_i ^{2}$ (see Eq. (\ref{int_eq_rho})). 
Solving the eigenvalue problem Eq. (\ref{int_eq_psi}) is equivalent to finding 
the Schmidt decomposition of the functions $\psi_x(x_1,x_2)$ and 
$\psi_y(y_1,y_2)$. 

 To this end we use the Mehler's formula \cite{serdelyi_1957},

\begin{equation}
\label{meheler}
e^{-(u^2+v^2)\frac{y^2}{1-y^2}+uv \frac{2y}{1-y^2}} = \sum_{l}^{\infty} 
\sqrt{1-y^2}\left(\frac{y}{2}\right) \frac{H_l(u) H_l(v)}{l!}\;,
\end{equation}

\noindent to write the Schmidt decomposition of Eqs. 
(\ref{function_q}) and (\ref{psi_y}),

\begin{equation}
\label{scmidt_app}
\psi(u,v) = \sum\limits_l^{\infty} \lambda_l\, \phi_l(u) \phi_l(v) \;, 
\end{equation}

Then $\psi_{x}(x_1,x_2)$ can be written as

\begin{equation}
\label{psi_x_sd}
\psi_x(x_1,x_2) = \sum\limits_l^{\infty} \lambda_l 
\lbrace \varphi_{l} \left( x_1 -\frac{x_0}{2} \right) \varphi_{l} \left( x_2 
+\frac{x_0}{2} \right) +
\varphi_{l} \left( x_1 +\frac{x_0}{2} \right) \varphi_{l} \left( x_2 
-\frac{x_0}{2} \right) 
\rbrace \; ,
\end{equation}

\noindent where $\varphi_{l}$ are the harmonic oscillator states,

\begin{equation}
\label{phi_oscillator_x}
\varphi_{l}(u) = \frac{ \left( \sqrt{2} 
\omega_x\right)^\frac{1}{8}}{\pi^\frac{1}{4} \sqrt{ 2^l l!}} 
e^{-\frac{\sqrt{\sqrt{2} \omega_x} u^2}{2}} H_l\left(\left( \sqrt{2} 
\omega_x\right)^\frac{1}{4} u\right) \; .
\end{equation}

Taking

\begin{equation}
\label{natural_orbitals}
\phi_{l}^{+}(u) = \frac{\varphi_{l} \left( u +\frac{x_0}{2} \right) + 
\varphi_{l} \left( u -\frac{x_0}{2} \right)}{\sqrt{2}} \;\;\;\; 
\mbox{and}\;\;\;\; \phi_{l}^{-}(u) = \frac{\varphi_{l} \left( u -\frac{x_0}{2} 
\right) - \varphi_{l} \left( u +\frac{x_0}{2} \right)}{\sqrt{2}}\; ,
\end{equation}

\noindent Eq. (\ref{psi_x_sd}) can be re-written as

\begin{equation}
\label{psi_x_sd_2}
\psi_x(x_1,x_2) = \sum\limits_l^{\infty} \lambda_l 
\lbrace \phi_{l}^{+} \left( x_1 \right)  \phi_{l}^{+} \left( x_2 \right) - 
\phi_{l}^{-} \left( x_1 \right) \phi_{l}^{-} \left( x_2 \right)
\rbrace \; .
\end{equation}

It is worth to mention that $ \left\langle \left. \phi_{l}^{+} \left( u \right) 
\right| \phi_{\tilde{l}}^{-} \left( u \right) \right\rangle = 0$, but  $ 
\left\langle \left. \phi_{l}^{+} \left( u \right) \right| \phi_{\tilde{l}}^{+} 
\left( u \right) \right\rangle = \delta_{l,\tilde{l}}$ and $ \left\langle \left. 
\phi_{l}^{-} \left( u \right) \right| \phi_{\tilde{l}}^{-} \left( u \right) 
\right\rangle = \delta_{l,\tilde{l}}$ only if the overlap $ \left\langle \left. 
\varphi_{l} \left( u - \frac{x_0}{2} \right) \right| \varphi_{\tilde{l}} \left( 
u + \frac{x_0}{2} \right) \right\rangle = 0$. 
This overlap decreases when $x_0$ increases and it vanish for $x_0 \to \infty$. 
Therefore, Eq. (\ref{psi_x_sd_2}) is the Schmidt decomposition of the wave 
function of Eq. (\ref{psi_x}) up from some interaction strength value large 
enough to guarantee that the minima $\pm x_0$ are sufficiently far from each 
other. Moreover, it is the Schmidt decomposition of the reduced density matrix 
 if $\lambda_l$ is replaced by $\Lambda_l=\lambda_l^2$. From Eq. 
(\ref{psi_x_sd_2}) it is clear that each occupancy $\Lambda_l$ is 
doubly degenerate with natural orbitals $\phi_{l}^{+} \left( u \right)$ and 
$\phi_{l}^{-} \left( u \right)$, then, the normalization is given by 
$\sum\limits_l^{\infty} \Lambda_l =1/2$.

Due to the separability of the wave function, the natural orbitals are the 
product of a natural orbital associated to $\psi_x(x_1,x_2)$ \textit{i.e.} 
$\phi_{l}^{+} \left( u \right)$ and $\phi_{l}^{-} \left( u \right)$, and a 
natural orbital associated to $\psi_y(y_1,y_2)$ with the following form,

\begin{equation}
\label{phi_oscillator_y}
\vartheta_{m}(v) = \frac{ \left( \varepsilon 
\sqrt{\varepsilon^2-1}\right)^\frac{1}{8}}{\pi^\frac{1}{4} \sqrt{ 2^m m!}} 
e^{-\frac{\sqrt{\varepsilon \sqrt{\varepsilon^2-1} } \;v^2}{2}} 
H_m\left(\sqrt{\varepsilon \sqrt{\varepsilon^2-1}} \;v\right) \; .
\end{equation}

After performing some algebra, we get the 
occupancies in the limit of large interaction strength parameter $g$ as

\begin{equation}
\label{sauval_2D}
\Lambda_{l,\tilde{l}} = \Lambda_{l}^{x} \, \Lambda_{\tilde{l}}^{y} \;,
\end{equation}

\noindent where

\begin{equation}
\label{sauval_2D_x}
\Lambda_{l}^{x}  = \frac{\left(1-\zeta\left( \omega_x\right)  \right)}{2 \left( 
1 + e^{-\frac{x_0^2\,\omega_x}{\sqrt{2}}} \right) } \zeta\left( 
\omega_x\right)^{l} \;\;\;\; \mbox{,}\;\;\;\; \zeta\left( \omega_x\right) = 
\left( \frac{ \left( 2 \omega_x^2\right) ^{\frac{1}{4}} - 1}{\left( 2 
\omega_x^2\right) ^{\frac{1}{4}} + 1}\right)^2 \;,
\end{equation}

\noindent and

\begin{equation}
\label{sauval_2D_y}
\Lambda_{\tilde{l}}^{y} = \left(1-\xi(\varepsilon)\right) 
\xi(\varepsilon)^{\tilde{l}} \;\;\;\; \mbox{,}\;\;\;\; \xi(\varepsilon) = 
\left( 
\frac{\left( \varepsilon^2 -1 \right) ^\frac{1}{4} - \sqrt{\varepsilon}}{\left( 
\varepsilon^2 -1 \right) ^\frac{1}{4} + \sqrt{\varepsilon}}\right)^2 \;.
\end{equation}


\section{Generalization to dimension higher than two}
\label{b}

Although all the calculations were heretofore carried out for two dimensions, 
the generalization to three and more dimensions is straightforward, we present 
here the generalization for dimension $D$. A system consisting of two particles 
in a $D$-dimensional anisotropic harmonic trap interacting via some potential 
which depends on the distance between particles, is given by the following 
Hamiltonian, 

\begin{equation}
\label{H_cal_D_an}
H = -\frac{1}{2}\left( \nabla_1^2+\nabla_2^2\right) + 
\frac{1}{2}\left\lbrace (x_{11}^2+x_{12}^2)+
\sum_{i=2}^{D}\varepsilon_{i-1}^2(x_{i1}^2+x_{i2}^2) \right\rbrace + g V\left( 
r_{12} ; \left\lbrace \gamma_{i} \right\rbrace \right)\; ,
\end{equation}

\noindent with $\varepsilon_{D-1} > \varepsilon_{D-2} > ... > 
\varepsilon_{1}>1$, and $x_{ij}$ denoting the $i$-th coordinate of the particle 
$j$. The minima satisfy 

\begin{equation}
\label{minD_V_eff}
\vec{r}_{min} =\left(\pm x_0,\ldots ,0 \right) \;\;\;\; \mbox{with $x_0>0$ 
given 
by}\;\;\;\; \frac{1}{2g}=-\left.\left( \frac{1}{r} \frac{\partial V}{\partial 
r}\right)\right\vert_{x_0} \,. 
\end{equation}

It is worth 
to notice that in dimension $D$ the minima of the potential in the isotropic 
case lie on the $D$-dimensional shell of radius $x_0$ (see Eq. 
(\ref{minD_V_eff})). 

The occupancies are

\begin{equation}
\label{auval_D}
\Lambda_{l_1,l_2,...,l_D} = \Lambda_{l_1}^{x} \, \prod_{i=2}^{D} 
\Lambda_{l_i}^{y} \;,
\end{equation}

\noindent where 
\begin{equation}
\label{sauval_2DD_x}
\Lambda_{l}^{x}  = \frac{\left(1-\zeta\left( \omega_x\right)  \right)}{2 \left( 
1 + e^{-\frac{x_0^2\,\omega_x}{\sqrt{2}}} \right) } \zeta\left( 
\omega_x\right)^{l} \;\;\;\; \mbox{,}\;\;\;\; \zeta\left( \omega_x\right) = 
\left( \frac{ \left( 2 \omega_x^2\right) ^{\frac{1}{4}} - 1}{\left( 2 
\omega_x^2\right) ^{\frac{1}{4}} + 1}\right)^2 \;,
\end{equation}

\noindent and

\begin{equation}
\label{sauval_2DD_y}
\Lambda_{\tilde{l}}^{y} = \left(1-\xi(\varepsilon)\right) 
\xi(\varepsilon)^{\tilde{l}} \;\;\;\; \mbox{,}\;\;\;\; \xi(\varepsilon) = \left( 
\frac{\left( \varepsilon^2 -1 \right) ^\frac{1}{4} - \sqrt{\varepsilon}}{\left( 
\varepsilon^2 -1 \right) ^\frac{1}{4} + \sqrt{\varepsilon}}\right)^2 \;.
\end{equation}

\noindent and the replacement $\varepsilon \mapsto \varepsilon_i$.

The linear entropy is

\begin{equation}
\label{le_final_D}
S_{L}^{D} = 1 - \frac{1}{2} \; \frac{1-\zeta(\omega_x)}{1+\zeta(\omega_x)} \; 
\prod_{i=2}^{D} \frac{1-\xi(\varepsilon_{i-1})}{1+\xi(\varepsilon_{i-1})} \;,
\end{equation}

\noindent with $\zeta(\omega_x)$ and $\xi(\varepsilon)$ as in Eqs. 
(\ref{sauval_2D_x}) and (\ref{sauval_2D_y}) respectively.

Finally, the R\'enyi entropies are given by

\begin{equation}
\label{S_VN_D}
S_{D}^{\alpha} = S_{x}^{\alpha}(\omega_x) + \sum_{i=2}^{D} 
S_y^{\alpha}(\varepsilon_{i-1})  \;,
\end{equation}

\noindent where 

\begin{equation}
\label{sRenyi_2D_x}
S^{\alpha}_{x} \left( \omega_x\right) = \frac{1}{1-\alpha} \log_2 \left(  
\frac{(1-\zeta\left( \omega_x\right))^ \alpha}{(1-\zeta\left( 
\omega_x\right)^\alpha)}
\right) + 1 \,,
\end{equation}

\noindent and

\begin{equation}
\label{sRenyi_2D_y}
S^{\alpha}_{y} (\varepsilon) = \frac{1}{1-\alpha} \log_2 \left(  
\frac{(1-\xi(\varepsilon))^ \alpha}{(1-\xi(\varepsilon)^\alpha)}
\right) \,.
\end{equation}

\noindent The von Neumann entropy can be obtained as a 
limiting case of the R\'enyi entropies when $\alpha\to 1$, with

\begin{equation}
\label{sSx}
S_{x}^{1}\left( \omega_x\right) = - \frac{\log_2\left( \left(1-\zeta\left( 
\omega_x\right) \right)^{(1-\zeta\left( \omega_x\right))} \zeta\left( 
\omega_x\right)^{\zeta\left( \omega_x\right)} \right)}{ \left(1-\zeta\left( 
\omega_x\right)\right)} + 1\;,
\end{equation}

\begin{equation}
\label{sSy}
S_{y}^{1}(\varepsilon) = - \frac{\log_2\left( 
\left(1-\xi(\varepsilon)\right)^{(1-\xi(\varepsilon))} 
\xi(\varepsilon)^{\xi(\varepsilon)} \right)}{ 
\left(1-\xi(\varepsilon)\right)}\;.
\end{equation}

\section{Inverse logarithmic interaction}
\addtocounter{figure}{-6}

\begin{figure}[h]
\begin{center}
\includegraphics[height=0.39\textwidth]{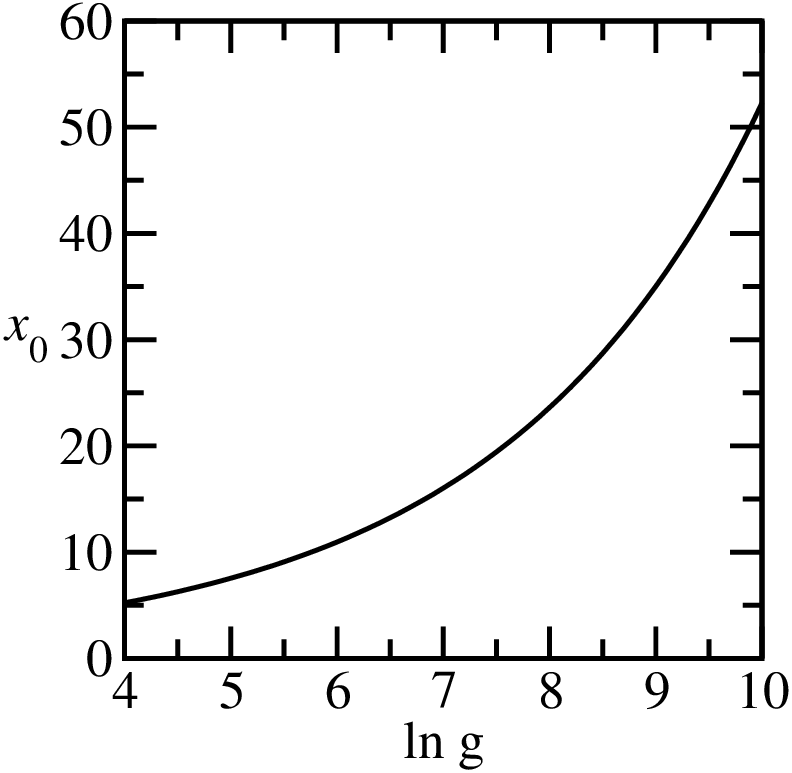}
\includegraphics[height=0.39\textwidth]{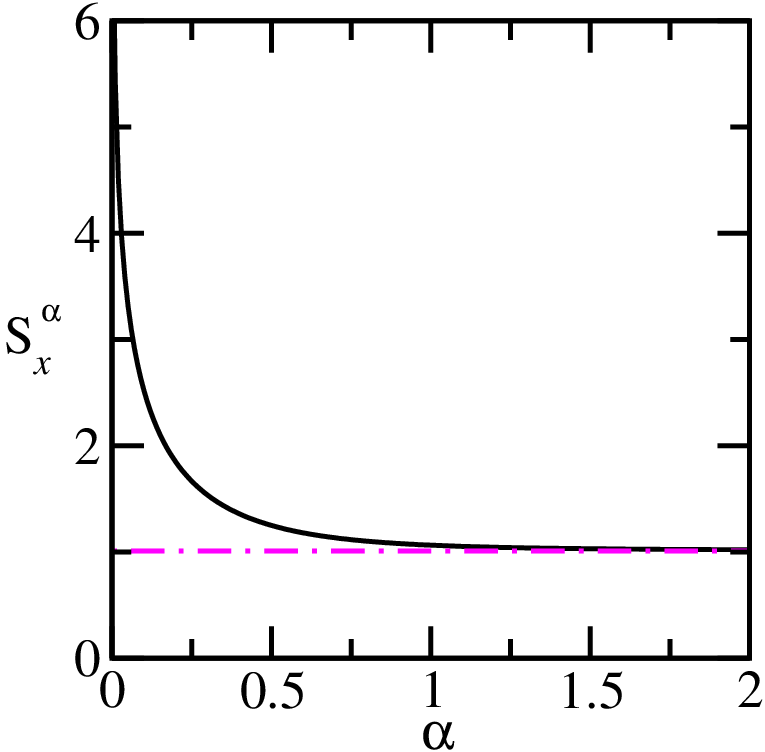}
\end{center}
\caption{  \label{f_S_inv_log} (a) Abscissa of the minima as a function of the interaction strength parameter $g$. (b) One-dimensional entropy terms $S_x^{\alpha}$ in the large interaction limit as a function of $\alpha$. The R\'enyi entropy diverges for $\alpha \to 0^{+}$ (max-entropy), and the limit $\alpha \to \infty$ (min-entropy) is depicted as a magenta dash-dotted line.
}
\end{figure}
\newpage

\end{document}